\documentclass[final,3p,times,twocolumn]{elsarticle}

\usepackage{graphicx}

\usepackage{amssymb}
\usepackage{amsmath}
\usepackage{textcomp}
\usepackage{url}

\journal{Journal of Quantitative Spectroscopy \& Radiative Transfer}

\begin{document}

\begin{frontmatter}



\title{ALIS: An efficient method to compute high spectral resolution polarized solar radiances using the Monte Carlo approach}


\author{Claudia Emde}
\ead{claudia.emde@lmu.de}
\author{Robert Buras}
\author{Bernhard Mayer}

\address{Meteorological Institute, Ludwig-Maximilians-University,
  Theresienstr. 37, D-80333 Munich, Germany}

\begin{abstract}
An efficient method to compute accurate polarized solar radiance spectra using
the (three-dimensional) Monte Carlo model MYSTIC has been
developed. Such high resolution spectra are
measured by various satellite instruments for remote sensing of atmospheric
trace gases. ALIS (Absorption Lines Importance Sampling) allows the
calculation of spectra by tracing photons at only one wavelength:
In order to take into account the spectral dependence of the
absorption coefficient a spectral absorption weight is calculated for
each photon path.
At each scattering event the local estimate method is combined with an
importance sampling method to take into account the spectral
dependence of the scattering coefficient.  Since each wavelength grid
point is computed based on the same set of random photon paths, the
statistical error is the almost same for all wavelengths and hence the
simulated spectrum is not noisy. The statistical error mainly results
in a small relative deviation which is independent of wavelength and can be
neglected for those remote sensing applications where differential
absorption features are of interest.

Two example applications are presented: The simulation of
shortwave-infrared polarized spectra as measured by GOSAT from which
CO$_2$ is retrieved, and the simulation of the differential optical
thickness in the visible spectral range which is derived from
SCIAMACHY measurements to retrieve NO$_2$. The computational speed of
ALIS (for one- or three-dimensional atmospheres) is of the order of or
even faster than that of one-dimensional discrete ordinate methods, in
particular when polarization is considered.
\end{abstract}

\begin{keyword}
radiative transfer \sep Monte Carlo method \sep polarization \sep
trace gas remote sensing \sep high spectral resolution  

\end{keyword}

\end{frontmatter}


\section{Introduction}
\label{sec:intro}

Monitoring of atmospheric trace gases is important to understand
atmospheric composition and global climate change. 
In particular climate models require information about the concentration
and global distribution of trace gases like e.g. H$_2$O, CO$_2$,
O$_3$, or CH$_4$. The
trace gases can be observed by measuring solar radiation which is
scattered and absorbed by the molecules. Several instruments have been
developed: satellite instruments provide global observations,
local measurements can be taken from the ground, from
air-plane, or from a balloon. Most instruments designed for
trace gas concentrations observations measure radiance spectra with
high spectral resolution. In the UV-Vis spectral range,
absorption of radiation is due to molecular transitions; at the same
time vibrational and rotational transitions can take place, which
results in band spectra where the individual absorption lines can not
be distinguished. Nevertheless each molecule type has its specific
absorption features so that the measured spectra include information
about the various trace gas concentrations. In the near-infrared
spectral range there are mainly vibrational transitions; here 
individual lines can be identified and used for trace gas
measurements.

Examples for currently operating satellite instruments that measure high
resolution radiance spectra of scattered solar radiation are SCIAMACHY on the
ENVISAT satellite (\citet{gottwald2006}), OMI on AURA (\citet{levelt2006}),
GOME-2 on METOP (\citet{callies2000}) and TANSO-FTS on GOSAT
(\citet{kuze2009}). SCIAMACHY and TANSO-FTS have the advantage of measuring
not only the radiance but also the polarization state of the radiation. While
extraterrestrial solar radiation is unpolarized, the radiance arriving
at the satellite is polarized due to
scattering by molecules, aerosols, or clouds and due to surface
reflection. The polarization information may therefore be used to reduce the
uncertainties in trace gas retrievals introduced by aerosols, clouds and
surface reflection.

The retrieval of trace gas concentrations from radiance spectra requires a
so called forward model, which can simulate such measurements for given
realistic atmospheric conditions. For the often used optimal
estimate retrieval method (\citet{rodgers2000}) it is important that the forward model is
fast because it has to be run several times until iteratively the atmospheric
composition is found which best matches the measured spectra.

A commonly used method to simulate solar radiative transfer is the
discrete ordinate method which was first described by
\citet{chandrasekhar50} and which has been implemented for instance
into the freely
available well known software DISORT (\citet{stamnes2000}). The DISORT
code however has the limitations that it assumes a plane-parallel
atmosphere (i.e. horizontal inhomogeneities can not be taken into
account) and that it neglects polarization. Polarization has been
included in the VDISORT code (\citet{schulz99}). 
The SCIATRAN code
(\citet{rozanov2005}) is also based on the discrete ordinate method.
It can optionally  take into account spherical geometry as well as
polarization (\citet{rozanov2006}).

Another method for the simulation of solar radiative transfer is the
Monte Carlo method (\citet{marchuk1980,marshak2005}), which is usually
much slower than the discrete ordinate
method. For this reason Monte Carlo methods have mostly been used for
simulations including inhomogeneous clouds (e.g. \citet{zinner2010}) for
which the plane-parallel approximation can not be applied. 
We have developed a new Monte Carlo method which
calculates high spectral resolution radiance spectra very efficiently. The
algorithm, named ALIS (Absorption Lines Importance Sampling),
does not require any approximations, in particular it can easily take
into account polarization and horizontal inhomogeneity. We show that
the computational
time of ALIS for high resolution radiance spectra is comparable to or
even faster than the discrete ordinate approach, especially if
polarization is included. This means that the algorithm is
sufficiently fast to be used as forward model for trace gas retrieval
algorithms. 
The basis of the ALIS method is that all wavelengths are
calculated at the same time based on the same random numbers.
This method which is sometimes called
``method of dependent sampling'' (\citet{marchuk1980}) has been used for various
applications, e.g. to calculate mean radiation fluxes in the near-IR
spectral range (\citet{titov1997}),  to compute multiple-scattering of
polarized radiation in circumstellar dust shells
(\citet{voshchinnikov1994}), or to calculate Jacobians (\citet{deutschmann2011}).
We have validated ALIS by comparison to the well-known and well-tested
DISORT program, originally developed and implemented by
\citet{stamnes2000} in FORTRAN77. We use a new version of the code
implemented in C (\citet{buras2011b}) with increased efficiency and
numerical accuracy.

\section{Methodology}

The new method Absorption Lines Importance Sampling (ALIS), which allows fast
calculations of spectra in high spectral resolution, has been
implemented into the radiative transfer model MYSTIC (Monte Carlo code for the
phYsically correct Tracing of photons In Cloudy atmospheres;
\citet{mayer2009}). MYSTIC is operated as one of several solvers of the
libRadtran radiative transfer package (\citet{mayer2005}). The common model
geometry of MYSTIC is a 3D plane-parallel atmosphere to simulate radiances or
irradiances in inhomogeneous cloudy conditions. The model can also be operated
in 1D spherical model geometry (\citet{emde2007}) which makes it suitable also
for limb sounding applications. Recently MYSTIC has been extended to handle
polarization due to scattering by randomly oriented particles, i.e. clouds,
aerosols, and molecules (\citet{emde2010}), and to handle topography
(\citet{mayer2010}). Several variance reduction
techniques were also introduced to MYSTIC in order to speed up the computation
of radiances in the presence of clouds and aerosols (\citet{buras2011}). 

\subsection{Monte Carlo method for solar atmospheric radiative transfer}

This section briefly describes the implementation of solar radiative
transfer in MYSTIC which is explained in detail in \citet{mayer2009}.
We describe only those details which are required to understand the
following sections about the ALIS method.

 In the
forward tracing mode ``photons''\footnote{We use the term "photon"
  to represent an imaginary discrete amount of electromagnetic energy
  transported in a certain direction. It is not related to the QED
  photon \cite{Mishchenko2009}.} are traced on their way through the
atmosphere.  The photons are generated at the top of the
atmosphere where their initial direction is given by the solar zenith
angle and the solar azimuth angle. 

Absorption and scattering are treated separately:
Absorption is considered by a photon weight according to Lambert-Beer's law:
\begin{equation}
  w_{\rm abs} = \exp \left(-\int_0^s \beta_{\rm abs}(s') {\rm d}s' \right)
  \label{eq:w_a}
\end{equation}
Here  ${\rm d}s'$ is a path element of the photon path and 
$\beta_{\rm abs} =\sum_{i=1}^N{\beta_{\rm abs,i}}$ is the total absorption
coefficient which is the sum of the $N$ individual absorption coefficients 
$\beta_{\rm abs,i}$ of molecules, aerosols, water droplets, and ice
crystals. The integration is performed over the full photon path.

The free path of a photon until a scattering
interaction takes place is sampled according to the probability
density function (PDF):
\begin{equation}
  P(s)= \beta_{\rm sca}(s) \exp \left({-\int_{0}^{s} \beta_{\rm sca}(s') {\rm d}s'}\right)
  \label{eq:pdf_s}
\end{equation}
Here $\beta_{\rm sca}=\sum_{i=1}^N{\beta_{\rm sca,i}}$ is the total
scattering coefficient of $N$ interacting particle and molecule types.

We use a random number $r\in[0,1]$ to decide which interaction takes
place. If there are $N$ types of particles and molecules at
the place of scattering, the photon interacts with the $j^{th}$ type
if the random number fulfills the following condition:
\begin{equation}
  \frac{\sum_{i=1}^{j-1} \beta_{\rm sca,i}}{\beta_{\rm sca}} < r \le
  \frac{\sum_{i=1}^{j} \beta_{\rm sca,i}}{\beta_{\rm sca}}
\end{equation}

At each scattering event the ``local estimate'' weight is calculated which corresponds to the
probability that the photon is scattered into the direction of the
detector and reaches it without further interactions:
\begin{equation}
  w_{{\rm le},is} =  \frac{ P_{\rm 11}(\theta_p)
  \exp\left(-\int{(\beta_{\rm abs}+\beta_{\rm sca}){\rm d}s'}\right)}{\cos(\theta_d)} 
\end{equation}
Here $\theta_p$ is the angle between photon direction (before scattering) and
the radiance direction.  The phase function P$_{\rm 11}$ (first
element of the scattering matrix) gives the probability
that the photon is scattered into the direction of the detector, ``$is$''
denotes the scattering order. In order to calculate the probability that the
photon actually reaches the detector the Lambert-Beer term for extinction $
\exp\left(-\int{(\beta_{\rm abs}+\beta_{\rm sca}){\rm d}s'}\right)$
needs to be included. Finally we have to divide
by the zenith angle of the detector direction $\theta_d$ to account for the
slant area in the definition of the radiance.  The contribution of the photon
to the radiance measured at the detector is then given as
\begin{equation}
  I_i=\sum_{is=1}^{\rm N_s} w_{{\rm abs},is} w_{{\rm le},is}
  \label{eq:rad_i}
\end{equation}
Here $i$ is the index of the photon, $N_s$ is the number of scattering events
along the photon path, and $w_{{\rm abs},is}$ is the absorption weight
(Eq.~\ref{eq:w_a}) evaluated at the scattering order $is$.  One can show
formally that the sum over the local estimate weights corresponds to a von
Neumann series which is a solution of the integral form of the radiative
transfer equation (see e.g. \citet{marshak2005}).

Additional weights are required to take into account polarization (\citet{emde2010})
and variance reduction techniques (\citet{buras2011}).

After tracing $N_p$ photons the radiance is given by the average
contribution of all photons: 
\begin{equation}
  I=\frac{\sum_1^{N_p} I_i}{N_p}
\end{equation}

The methods described above are implemented for monochromatic radiative
transfer. If one wants to calculate a radiance spectrum using these
methods one has to calculate
all spectral points subsequently. Here usually a very high accuracy is
required in order to distinguish spectral features from
statistical noise which means that such calculations are
computationally expensive. 

\subsection{Calculation of high spectral resolution clear-sky radiance spectra}

\begin{figure}[h]
  \centering
  \includegraphics[width=0.48\textwidth]{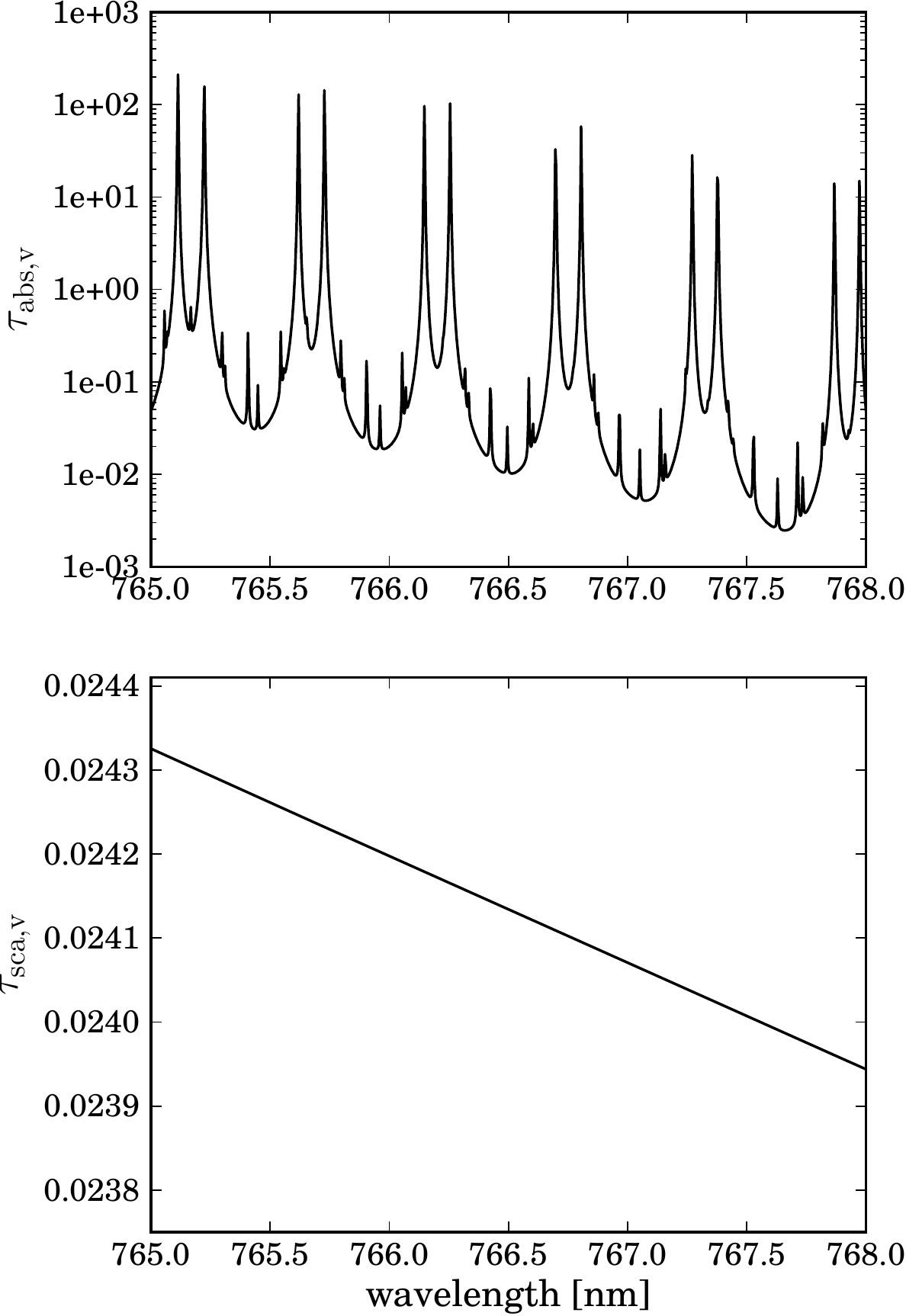}
  \caption{Integrated vertical optical thickness of molecular absorption (top)
    and molecular scattering (bottom).}
  \label{fig:optdepth}
\end{figure}

In the following an efficient method how to compute high spectral
resolution radiance spectra will
be described and demonstrated on the example of the spectral region from
765--768~nm in the O$_2$A absorption
band where we calculate the spectrum with a resolution of
0.003~nm. The line-by-line gas absorption coefficients have been computed using the
ARTS model (\citet{buehler2005}, \citet{eriksson2011}) for the standard mid-latitude
summer atmosphere (\citet{Anderson1986}).
Fig.~\ref{fig:optdepth} shows the vertically integrated optical thickness of molecular
absorption $\tau_{\rm abs,v}$ (top) and scattering $\tau_{\rm sca,v}$
(bottom). 
Whereas the scattering optical thickness for the cloudless,
aerosol-free atmosphere is rather small and almost constant,
it varies only from about 0.0239 to 0.0243, the absorption
optical thickness varies over five orders of magnitude, from about
10$^{-3}$ to 10$^{2}$ (note the logarithmic scale).

As mentioned in the previous section, absorption is considered
separately by calculating the absorption weight $w_{\rm abs}$
(Eq.~\ref{eq:w_a}). In order to calculate a radiance spectrum taking
into account the spectral variation of the absorption coefficient
$\beta_{\rm abs}$ we can easily calculate the absorption weight for each
wavelength and get a spectrally dependent absorption weight vector:
\begin{equation}
  \label{eq:w_abs_lam}
  w_{\rm abs}(\lambda) = \exp \left(-\int_0^s \beta_{\rm abs}(\lambda,
    s') {\rm d}s' \right)
\end{equation}
Here $\lambda$ denotes the wavelength of the radiation.
In practice the integral corresponding to the absorption optical
thickness $\tau_{\rm abs}=\int_0^s \beta_{\rm abs} {\rm d}s'$ is
calculated step by step while the photon 
moves through the layers/boxes of the discretized model
atmosphere (see \citet{mayer2009}):
\begin{equation}
  \tau_{\rm abs}(\lambda)=\sum_{p}{\beta_{\rm abs}(\lambda, p) 
    {\rm \Delta s}_p}
\end{equation}
Here the $p$ denotes the step index along the photon path, 
and ${\rm \Delta s}_p$ is the pathlength of step $p$. 
We also include the spectrally dependent absorption coefficient
$\beta_{\rm abs}(\lambda)$ in the local estimate weight 
$w_{{\rm  le},is}(\lambda)$.
Thus we only need to trace the photons for one wavelength,
calculate the spectral absorption weights
and get the full radiances spectrum with high spectral resolution. For
each photon we get (compare Eq.~\ref{eq:rad_i}): 
\begin{equation}
  I_i (\lambda)= \sum_{is=1}^{\rm N_s} w_{{\rm abs},is}(\lambda) w_{{\rm le},is}(\lambda)
\end{equation}
\begin{figure}[h]
  \centering
  \includegraphics[width=0.48\textwidth]{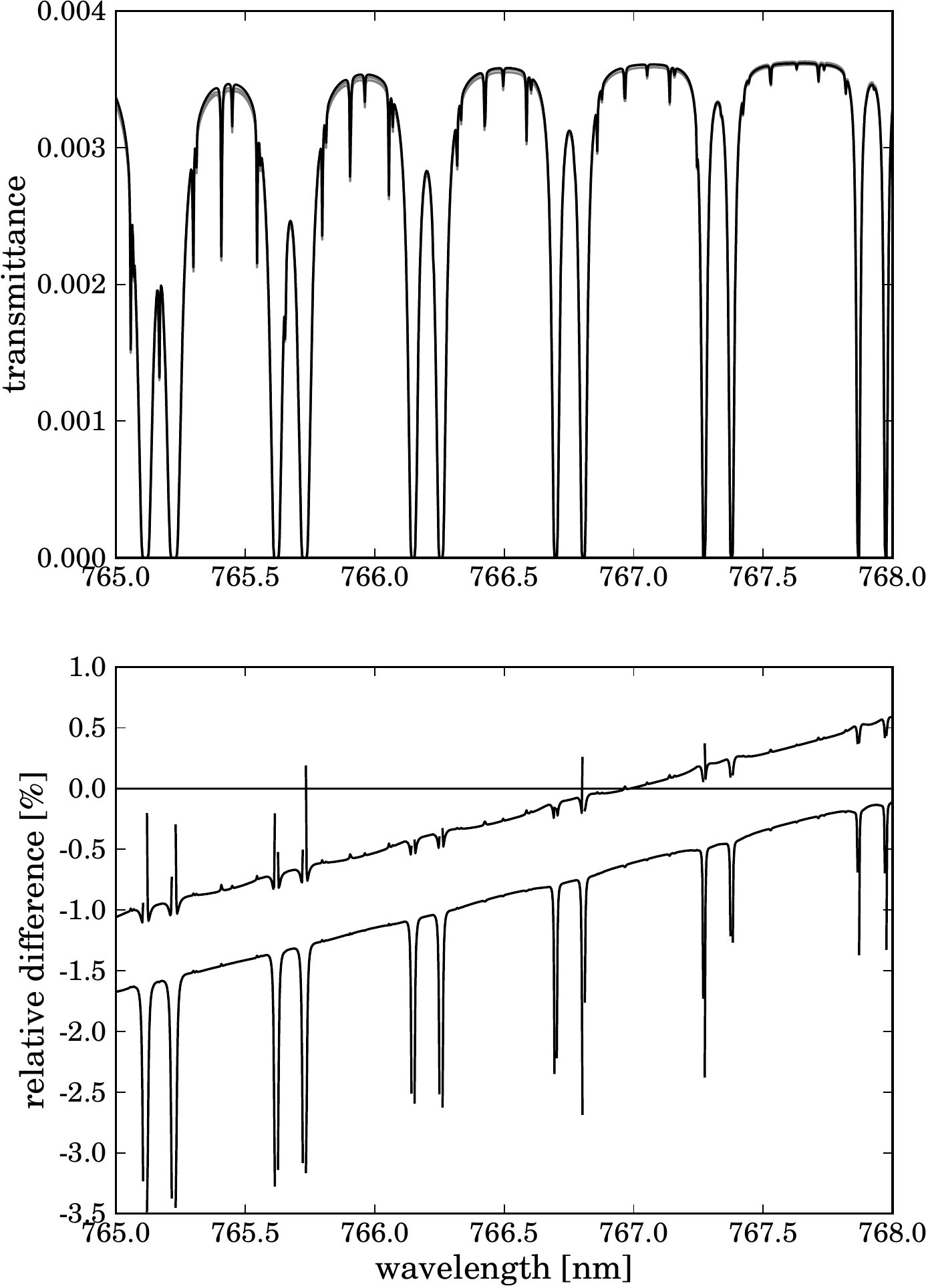}
  \caption{Radiance spectra calculated with MYSTIC in comparison to
    DISORT calculations. The top panel shows the transmittance
    (radiance normalized to extraterrestrial irradiance) spectra of
    two independent MYSTIC runs (grey lines) and the
    DISORT result (black line) and the
    bottom panel shows the relative differences between the MYSTIC runs
    and DISORT in percent.}
  \label{fig:only_abscorr}
\end{figure}
Fig.~\ref{fig:only_abscorr} shows two spectral calculations using this
method. 
Here we assumed that the sun is in the zenith and the sensor is on the
ground and measures with a viewing angle of 60$^\circ$. We did not
include any sensor response function. The upper
panel shows the transmittance spectra (radiance divided by
extraterrestrial irradiance) and the lower panel shows the
relative difference to the DISORT solver operated
with 32~streams. 
The MYSTIC calculations with 10$^6$ photons took 13~s on a single 
processor with 2~GHz CPU (all computational times in the following
refer to this machine),  the DISORT calculation took 25~s.
The relative difference between MYSTIC and DISORT is less than about
2\% with some exceptions where the transmittance is almost zero.
The spectral features in the MYSTIC calculations are well
resolved.
The two MYSTIC runs used exactly the same setup but the results show
a deviation between each other and with respect to the DISORT result.
This deviation is due to the statistical error of the Monte Carlo
calculation, with 10$^6$ photons the standard deviation is 0.66\%.
Hence the deviation can be reduced by running more photons. 
Since the same photon paths
are used to compute all wavelengths the deviation is the same at all
spectral data points and the spectra are not noisy. 
However the deviation shows a spectral dependence which is not a
statistical error but can be attributed to
the spectral dependence of Rayleigh scattering which has been
neglected so far. In the calculation $\beta_{\rm sca}$ was
set to a constant value corresponding to $\beta_{\rm sca}$ at
766.5~nm.
In the next section we will describe how to include
the spectral dependence of the scattering coefficient.  

\subsection{Importance sampling for molecular scattering}

Eq.~\ref{eq:pdf_s} is the PDF which we use
for sampling the free pathlength of the photons, where the scattering coefficient
$\beta_{\rm sca}$ now becomes wavelength dependent. We want to use
this PDF for sampling the pathlength for all wavelengths. In order to
ensure that the results are correct for all wavelength we 
introduce a correction weight (importance sampling method, see
e.g. \citet{ripley2006}):
\begin{align}
    &w_{\rm sca1}(\lambda, s) \\  
  &= \frac{\beta_{\rm sca}(\lambda, s) \exp \left(-\int_0^s \beta_{\rm sca}(\lambda, s') {\rm d}s'
    \right)} {\beta_{\rm sca}(\lambda_c, s) \exp \left(-\int_0^s \beta_{\rm sca}(\lambda_c, s') {\rm d}s'
    \right)} \nonumber  \\
  &= \frac{\beta_{\rm sca}(\lambda, s)}{\beta_{\rm sca}(\lambda_c,
    s)} \exp \left(-\int_0^s (\beta_{\rm sca}(\lambda, s')- \beta_{\rm
      sca}(\lambda_c, s')){\rm d}s' \right) \nonumber     
\end{align}
Here $\lambda_c$ ($c$ for ``computational'') is the wavelength
corresponding to the scattering
coefficient that is used to sample the photon free path.
As in the previous
section we write the second part of this expression as a sum over the model layers/boxes:
\begin{align}
  \label{eq:w_sca1}
  w_{\rm sca1}(\lambda,s) &= 
  \frac{\beta_{\rm sca}(\lambda, s)}{\beta_{\rm sca}(\lambda_c, s)}
   \\\nonumber 
  &\times \exp\left(-\sum_{p}{ (\beta_{\rm sca}(\lambda, p) - 
      \beta_{\rm sca}(\lambda_c, p)) \Delta s_p }\right)
\end{align}
The probability that the photon is scattered into a direction with
scattering angle
$\theta_p$ is given by the phase function $P_{11}(\lambda, \theta_p)$.
So we need
another weight to correct for the spectral dependence of the phase
function which can again easily be derived using the importance
sampling method:
\begin{equation}
  \label{eq:w_sca2}
  w_{{\rm sca2},is}(\lambda, s)=\frac{
    P_{11}(\lambda  ,\theta_p, s)
  }{P_{11}(\lambda_c, \theta_p, s)}
\end{equation}
Here $s$ is the location at which the photon is scattered.
Note that in the case where we have only molecules as interacting
particles and neglect depolarization $P_{11}$ is the Rayleigh phase function 
\begin{equation}
  P_{11} (\theta_p)=\frac{3}{4} (1+\cos^2 \theta_p)
\end{equation}
Also, as long as we neglect the wavelength dependence of the
Rayleigh depolarization factor (see e.g. \citet{hansen1974})
the Rayleigh phase function is wavelength-independent and 
$w_{{\rm sca2},is}(\lambda)=1$.

The final result for the contribution of a photon including the
spectral dependence of absorption and scattering to the
spectral radiance is:
\begin{align}
 & I_i (\lambda) = \\ \nonumber
 & \sum_{\rm is=1}^{\rm N_s} w_{{\rm abs},is}(\lambda) 
  \left(\prod_{is'=1}^{is} w_{{\rm sca1},is'} (\lambda, s')  w_{{\rm
        sca2},is'} (\lambda,s') \right)
  w_{{\rm le},is}(\lambda)
\end{align}
\begin{figure*}[t!]
  \centering
  \includegraphics[width=.8\textwidth]{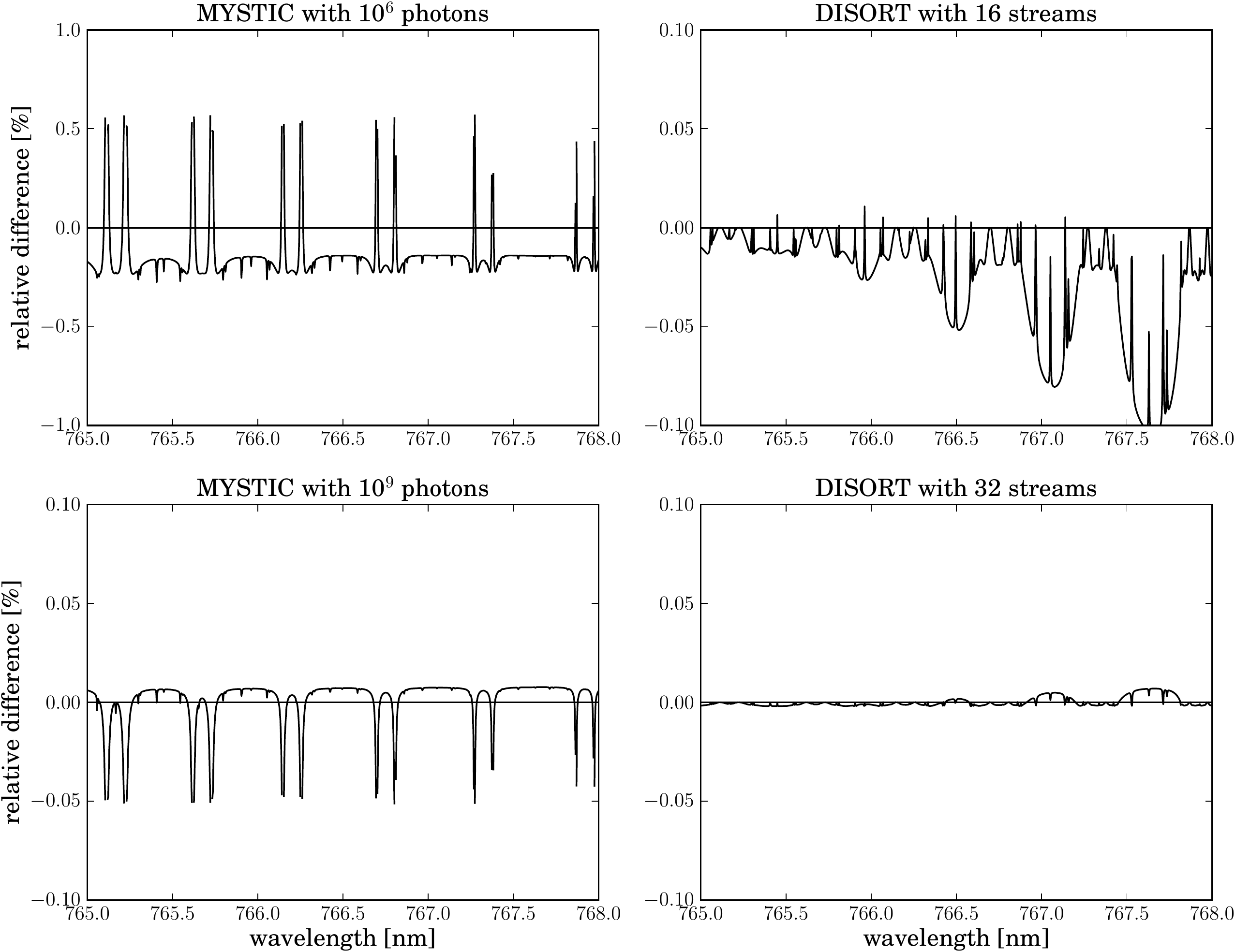}
  \caption{Relative differences of various model setups with respect
    to a DISORT calculation with 64 streams. The left panels show MYSTIC
    calculations with 10$^6$ and 10$^9$ photons respectively. The
    right panels show DISORT calculations with 16 and 32 streams respectively.}
  \label{fig:scacorr}
\end{figure*}

Now we calculate again the spectrum in the O$_2$A-band from 765--768~nm with
$\lambda_c=$765~nm and compare the result with an
accurate DISORT calculation using 64~streams. The top left panel of
Fig.~\ref{fig:scacorr} shows the relative difference of a MYSTIC run with
10$^6$ photons, which takes 14.6~s, to DISORT. We see that there is still a
relative deviation of about 0.4\% which is due to the statistical error of the Monte Carlo
calculation, but the spectral dependence of the deviation is now removed because
we have corrected the spectral dependence of the scattering coefficient. In
order to check whether the method is correctly implemented without any bias
(apart from the statistical error) we performed a MYSTIC run with 10$^9$
photons. The result is shown in the lower left panel. The spectrally independent
deviation has almost vanished ($<$0.01\%)
and the relative difference between MYSTIC and DISORT is below
0.05\%. For
comparison we show in the right panels of the figure DISORT runs with 16 and
32 streams respectively compared to the DISORT run with 64 streams. The
difference between DISORT (16 streams) and DISORT (64 streams) is actually
larger than the difference between MYSTIC (10$^9$ photons) and DISORT (64
streams).

It should be noted that this Monte Carlo method does only work well as long as the
scattering coefficient does not vary too much within the simulated wavelength
region. Else the scattering weights can obtain values very far from 1,
resulting in large statistical noise and slow convergence.

\subsection{Calculation of high resolution spectra including  aerosol and
  cloud scattering}

It is straightforward to apply the method to an atmosphere including
clouds and/or aerosols. We just need to use the total absorption and scattering coefficients
\begin{eqnarray}
  \beta_{\rm abs}(\lambda) =\sum_{i=1}^N{\beta_{\rm abs,i}}(\lambda) \\
  \beta_{\rm sca}(\lambda) =\sum_{i=1}^N{\beta_{\rm sca,i}}(\lambda)
\end{eqnarray}
and the average phase function given by
\begin{equation}
  P_{\rm 11}(\lambda)=\frac{\sum_{i=1}^N{\beta_{\rm sca,i}(\lambda)
      P_{\rm 11,i}(\lambda)}}
  {\beta_{\rm sca}(\lambda)}
\end{equation}
Here $N$ is the number of interacting particles/molecules. These
quantities can be used to compute the wavelength dependent weights
$w_{\rm abs}(\lambda)$ (Eq.~\ref{eq:w_abs_lam}), $w_{\rm
  sca1}(\lambda)$ (Eq.~\ref{eq:w_sca1}) and $w_{\rm   sca2}(\lambda)$
(Eq.~\ref{eq:w_sca2}).
In MYSTIC we so far consider only the spectral dependence of molecular
scattering because the spectral
dependence of cloud and aerosol
scattering can safely be neglected in narrow wavelength intervals.

\section{Applications}

\subsection{Simulation of polarized near infrared spectra in cloudless conditions}

The Greenhouse Gases Observing Satellite (GOSAT) determines the
concentrations of carbon dioxide and methane globally from space.
The spacecraft was launched 
on January 23, 2009, and has been operating properly since then. 
Information about the project can be found on the web-page
\url{http://www.gosat.nies.go.jp}.
GOSAT carries the Thermal and Near Infrared Sensor for Carbon
Observation Fourier-Transform Spectrometer (TANSO-FTS)
(\citet{kuze2009}) which measures
in 4 spectral bands (band~1: 0.758--0.775~$\mu$m, band~2: 1.56--1.72~$\mu$m,
band~3: 1.92--2.08~$\mu$m, band~4: 5.56--14.3~$\mu$m). The
spectral resolution in all bands is 0.2~cm$^{-1}$. For the visible
spectral range (bands 1--3) polarized observations are performed. In
order to analyze this data a fast polarized radiative transfer code is
required. The Monte Carlo approach which is described in this
study is an alternative to commonly used discrete ordinate or
doubling and adding codes. The approach is fully compatible to the
implementation of polarization in MYSTIC as described in
\citet{emde2010} and validated in \citet{kokhanovsky2010},
because the weight vector which is calculated to
take into account
the polarization state of the photon does not interfere with the
spectral weights. An advantage of the Monte Carlo approach
is of course that it is easy to take into account horizontal
inhomogeneities of clouds, aerosols, and molecules. 

In the following we show an example simulation where we selected a
spectral window of band~3 from 4815--4819~cm$^{-1}$ (corresponding to
$\approx$2.075--2.077~$\mu$m) in the near
infrared. The radiance simulation is performed with a spectral
resolution of 0.01~cm$^{-1}$. The atmospheric profiles (pressure,
temperature, trace gases) correspond to the standard mid-latitude
summer atmosphere as defined by \citet{Anderson1986}.
The molecular absorption coefficients
have been computed using the ARTS line-by-line model. We assume 
a thin maritime aerosol layer with an optical thickness of
0.05 at 2~$\mu$m. We took the refractive indices and the size distribution data
from \citet{hess1998} (maritime clean aerosol mixture)
and performed Mie calculations to obtain the
aerosol optical properties including the phase matrix.  
We assume an underlying water surface
which is modeled using the reflectance matrix as defined in
\citet{mishchenko1997}. The reflectance matrix
is based on the Fresnel equations, on \citet{cox54a,cox54b} to
describe the wind-speed dependent slope of the waves, and on
\citet{tsang1985} to account for shadowing effects. The wind speed was taken
to be 5m/s. The viewing angle of the FTS is 30$^\circ$ and simulations have been
performed for the principal plane assuming different solar zenith
angles $\theta_0$. The full Stokes vector has been computed for all setups.

\begin{figure}[h]
  \centering
  \includegraphics[width=0.48\textwidth]{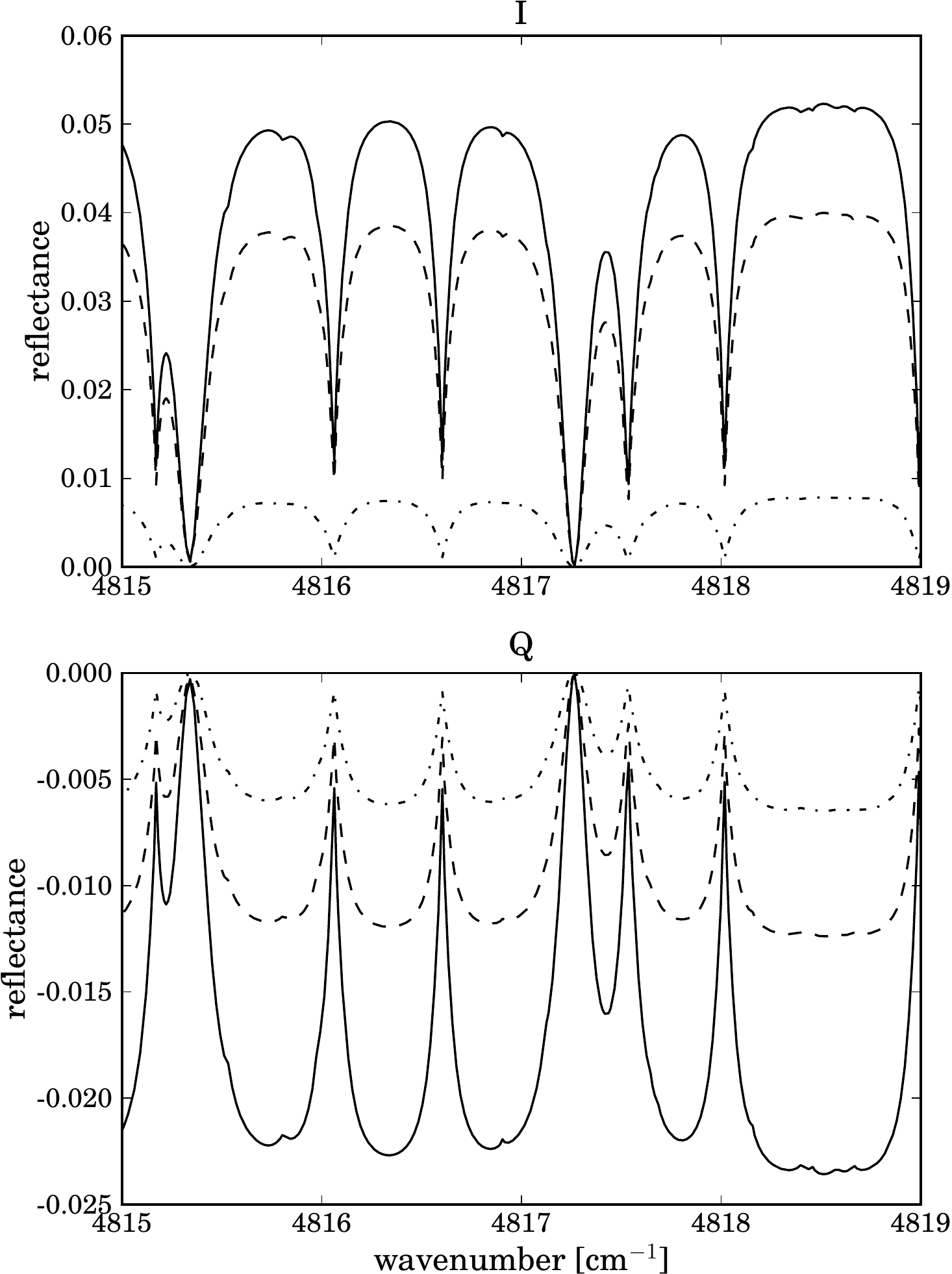}
  \caption{Simulated GOSAT spectra over ocean. The assumed wind
    speed is 5~m/s. The viewing angle of the FTS is 30$^\circ$. The
    line-styles correspond to different solar zenith angles, the
    solid line corresponds to 30$^\circ$, i.e. the sun glint is observed,
    the dashed and the dash-dotted lines correspond to 20$^\circ$ and
    60$^\circ$, respectively. All simulations are in the principal
    plane. The upper panel shows the normalized reflected
    intensity $I$ and the lower panel shows the polarization
    difference $Q$.}
  \label{fig:gosat}
\end{figure}
Fig.~\ref{fig:gosat} shows the simulated GOSAT spectra. The solid
lines correspond to $\theta_0$=30$^\circ$, in this case
the FTS observes the center of the sun glint, therefore this spectrum
shows the highest reflectance. The dashed lines are for
$\theta_0$=20$^\circ$, still in the sun-glint region and
the dashed-dotted lines are for $\theta_0$=60$^\circ$ which
is not influenced much by the sun glint. The computation time for
each polarized spectrum using 10$^6$
photons was 2~minutes and 25~seconds,
the standard deviation (approximately the same for each Stokes vector
component) for $\theta_0$=20$^\circ$ and $\theta_0$=30$^\circ$ is
0.03\%, for $\theta_0$=60$^\circ$ it is
0.16\%.

\subsection{Simulations of differential optical thickness in broken cloud conditions}

Retrievals of the tropospheric NO$_2$ column from SCIAMACHY data are based on the
differential optical absorption spectroscopy (DOAS) method
(\citet{richter2002, richter2005}). For this
method the measured spectra are converted to so-called
differential optical thicknesses defined as
\begin{equation}
  D(\lambda)=\ln( I_{\rm TOA}(\lambda)) - P_3(\lambda)
\end{equation}
where $I_{\rm TOA}(\lambda)$ is the reflectance at the top of the
atmosphere. $P_3(\lambda)$ is a third degree least square
polynomial fit of the logarithm of $I_{\rm TOA}(\lambda)$ with respect
to the wavelength, which removes the slowly varying part of the
spectrum. The conversion of $I_{\rm TOA}(\lambda)$ into $D(\lambda)$
improves the contrast of the NO$_2$ absorption line depths and thereby
the accuracy of the retrieval. The retrieval algorithm minimizes the
function 
\begin{equation}
  F(\lambda, V_{\rm NO_2, ret}) = \left| D(\lambda, V_{\rm NO_2, true})
    - D(\lambda, V_{\rm NO_2, ret}) \right|
\end{equation}
where $V_{\rm NO_2, true}$ and $V_{\rm NO_2, ret}$ are the true and the
retrieved tropospheric NO$_2$ columns, respectively.

\begin{figure}[h]
  \centering
  \includegraphics[width=.48\textwidth]{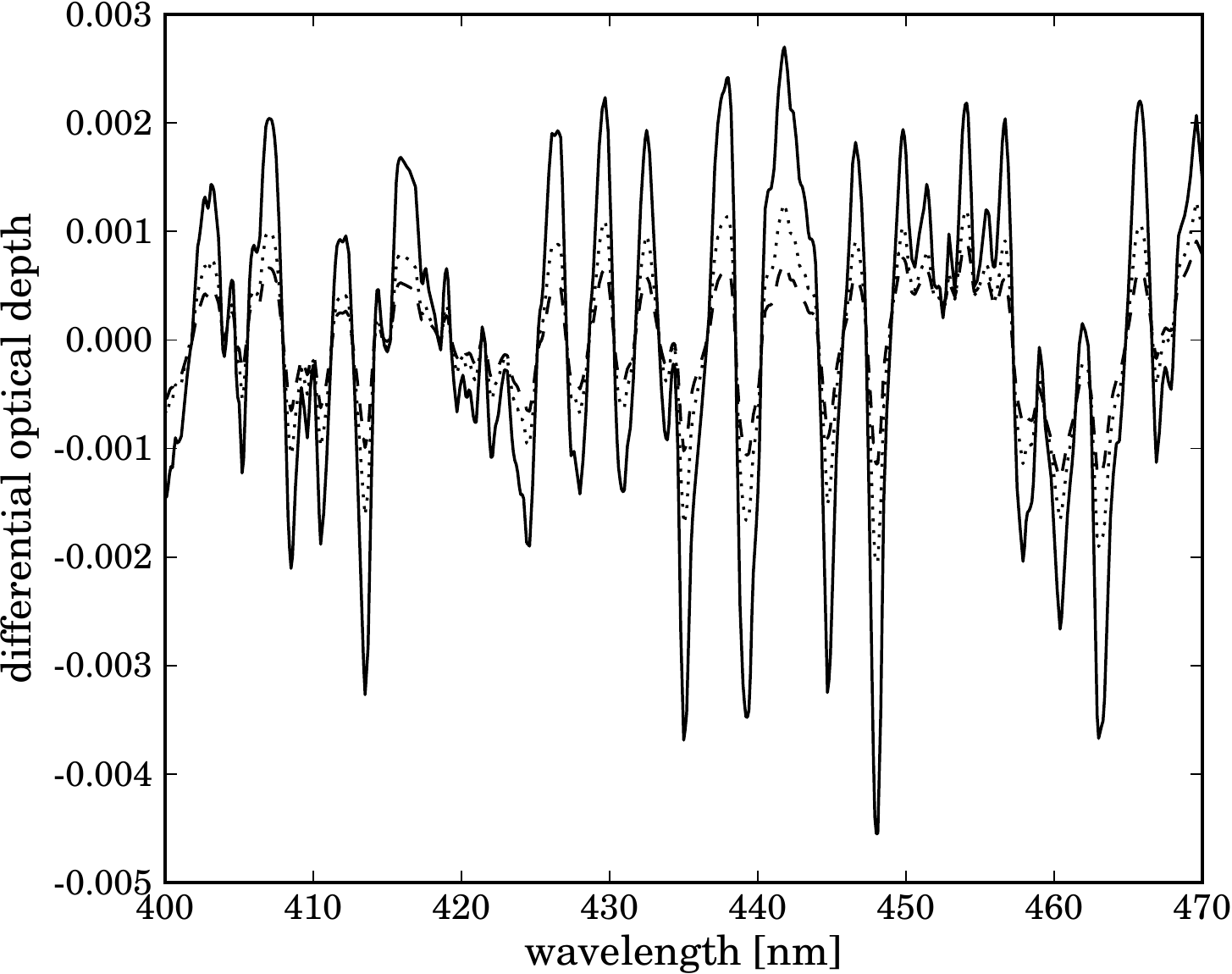}
  \caption{Differential optical thickness calculated for three
    different NO$_2$ profiles, corresponding to low (-~-), medium ($\cdots$)
    and highly polluted (---) conditions. The spectra have been
    computed using MYSTIC with 10$^5$ photons.}
  \label{fig:doas}
\end{figure}
Our new method allows efficient computations of $D(\lambda)$. 
As an example Fig.~\ref{fig:doas} shows spectra for
three different NO$_2$ profiles, corresponding to low, medium and
highly polluted conditions. The Lambertian surface albedo was set to
0.1 and the solar zenith angle to 32$^\circ$. 
NO$_2$ and O$_3$ profiles are the same
as used in the study by \citet{vidot2010}. 
The NO$_2$ absorption cross sections have been taken from
\citet{Burrows1998}, ozone absorption was also included in the
simulations using the cross sections by \citet{Molina1986}.
The spectral resolution of the simulation is 0.1~nm.
%
%

\begin{figure*}[t!]
  \centering
  \includegraphics[width=1.\textwidth]{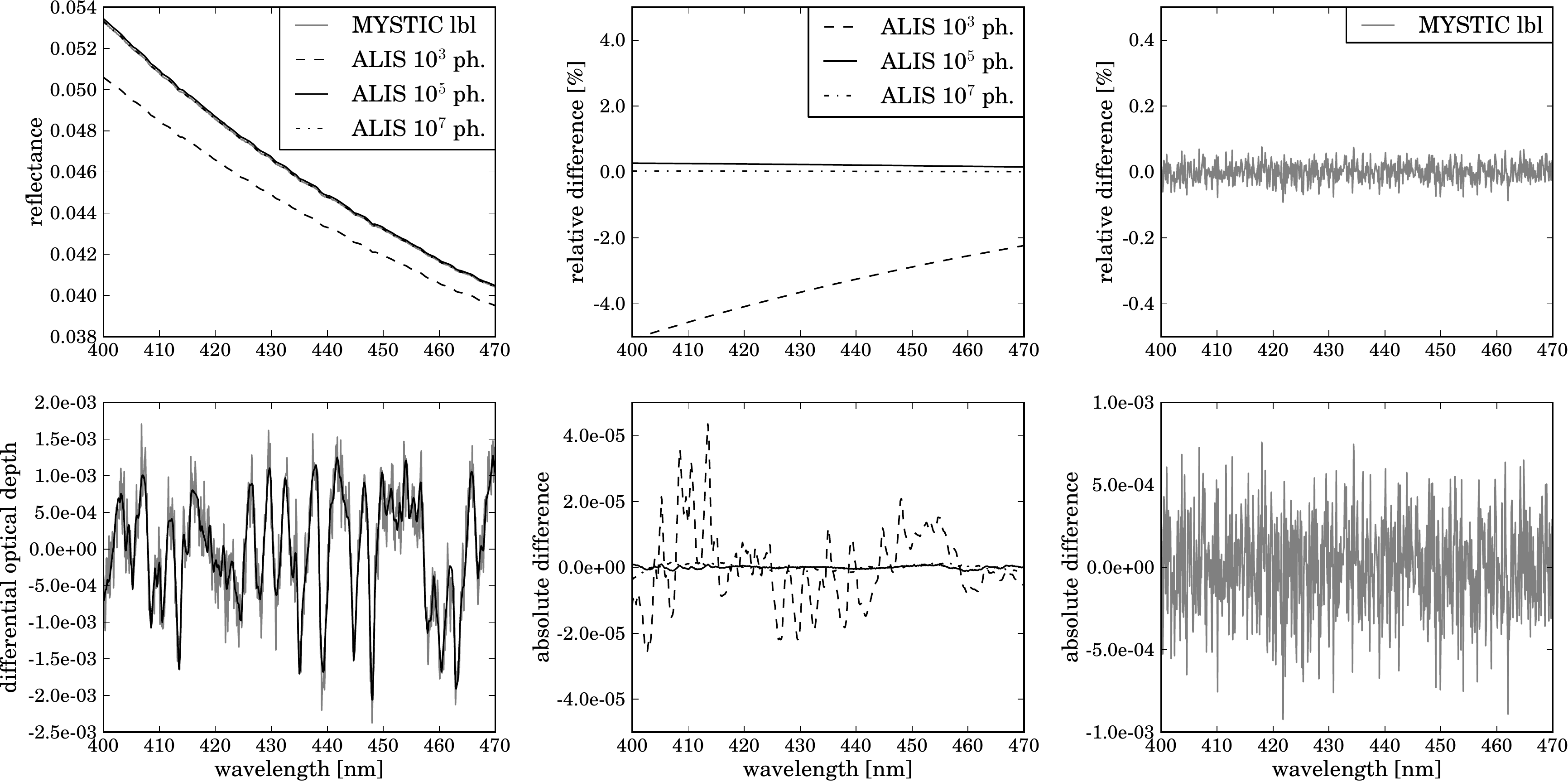}
  \caption{Simulations for NO$_2$ profile corresponding to
    medium polluted conditions. The top left panel shows the reflectance and
    the bottom left panel the differential optical thickness. The black lines
    correspond to Monte Carlo simulations using the ALIS method with
    different number of photons (10$^3$, 10$^5$, and 10$^7$), the
    grey line shows a Monte Carlo simulation without ALIS (all wavelengths
    are calculated independently). The middle panels show differences
    w.r.t. DISORT for the ALIS simulations and the right plots show
    differences w.r.t. DISORT for the MYSTIC calculation without using ALIS.}
  \label{fig:alis_vs_lbl}
\end{figure*}
Fig.~\ref{fig:alis_vs_lbl} shows calculations for the NO$_2$ profile
corresponding to medium polluted conditions. The top left panel shows the
reflectance, where the grey line corresponds to a MYSTIC calculation
without using ALIS, i.e. all wavelengths are simulated subsequently
using 10$^7$ photons for each wavelength. The standard deviation for
each wavelength is about 0.03\%. This calculation took
33~h~12~min on one CPU. The black line shows the calculations
using ALIS with different numbers of photons. The calculation using
10$^3$ photons took 0.9~s, the one with  10$^5$ photons took 38~s, and
the one with 10$^7$ photons took 63~min~53~s.
Obviously the absorption features of NO$_2$ are barely visible in the
reflectance plot. There is a deviation between the simulations
which is due to the
statistical error of the simulation using ALIS with 10$^3$ or 10$^5$
photons. The top middle panel shows the relative differences of the
ALIS simulations w.r.t. DISORT, which requires 30~s computation time
with 32 streams. Obviously the relative difference decreases with
increasing number of photons. The top right panel shows the relative
difference between the MYSTIC calculation without ALIS and DISORT. The
relative difference is less than 0.1\% and shows the typical Monte
Carlo noise. 

The bottom left panel shows the differential optical thicknesses
derived from the 
simulations. Here the statistical noise of the MYSTIC calculation
without ALIS (grey line) is clearly visible. All ALIS simulations
result in a smooth differential optical thickness because all wavelengths are calculated
based on the same photon paths. This yields a relative deviation in the
reflectance which is independent of wavelengths and can be removed 
completely by subtracting the fitted polynomial. 
The bottom middle panel shows the absolute difference between the
differential optical thicknesses derived from the ALIS simulations and
the one derived from the DISORT simulation. Even for
the simulation with only 10$^3$ photons the differential optical
thickness is quite accurate, the difference w.r.t. DISORT is of the order of a few
per-cent. Using 10$^5$ photons or more yields very accurate
differential optical thicknesses, the difference is here well below
1\%. The bottom right panel shows the difference between
the MYSTIC calculation
without ALIS and the DISORT calculation. The difference is of the same
order of magnitude as the differential optical thickness itself and
hence the accuracy of this simulation would not be sufficient for
NO$_2$ retrievals.

Fig.~\ref{fig:alis_vs_lbl} clearly demonstrates that a common Monte
Carlo approach which calculates wavelength by wavelength
sequentially is extremely inefficient for simulations of the
differential optical thickness because each wavelength has an
independent statistical error which is larger than the
absorption features unless a huge number of photons is used.
In order to obtain a result with an accuracy
comparable to ALIS with 10$^3$ photons (0.9~s computation time),
MYSTIC without ALIS would require at least 10$^9$ photons per wavelength
which would take about 138~days computation time.
\begin{figure}[h]
  \centering
  \includegraphics[width=.48\textwidth]{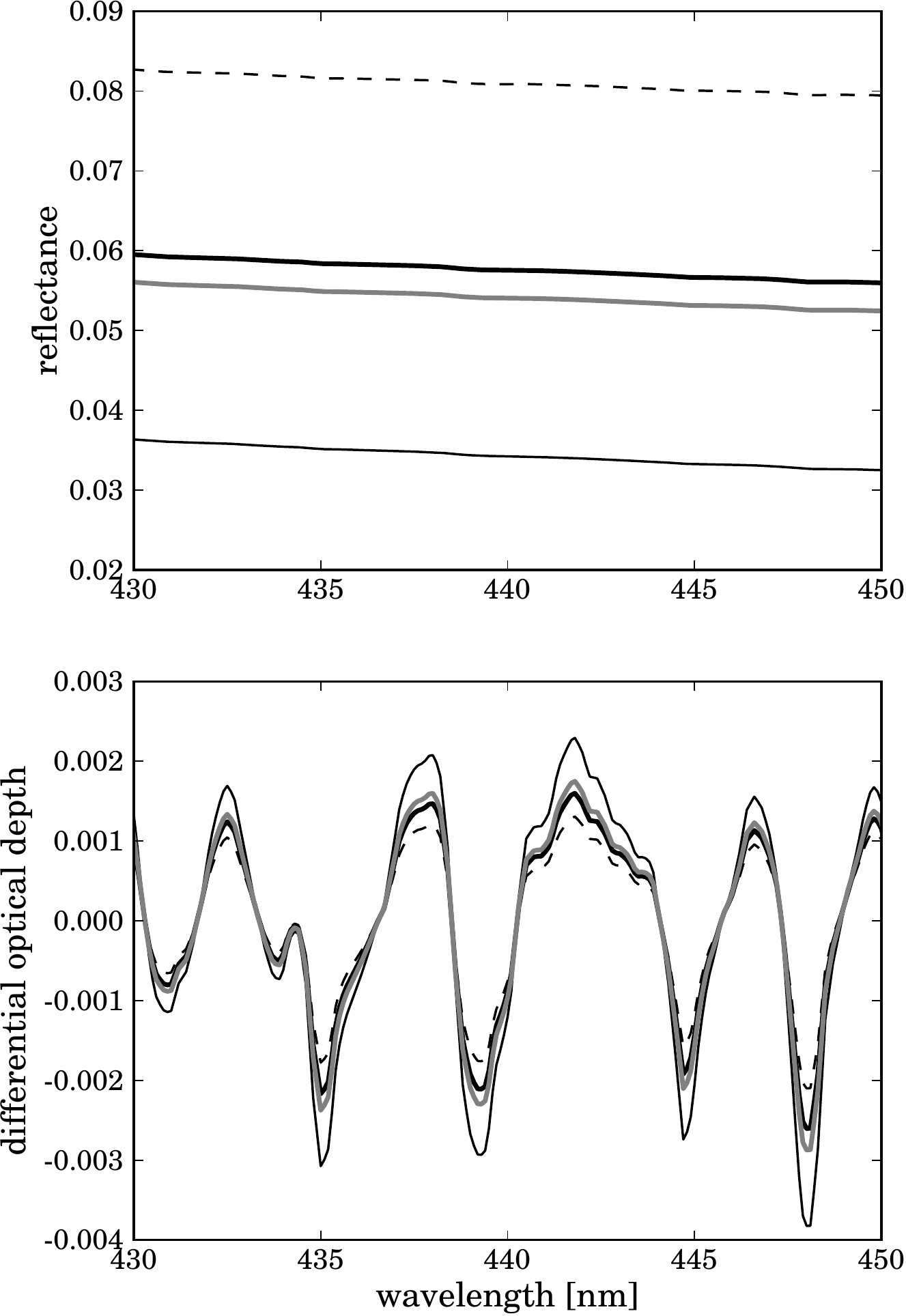}
  \caption{Impact of cirrus clouds on a spectrum used for NO$_2$
    retrievals. The cirrus clouds are modeled as 1$\times$1$\times$1~km$^3$
    cubes, the cloud fraction in the model domain is 0.5. Black lines
    correspond to the independent pixel (IPA) calculation and grey lines to
    the 3D calculation. Upper panel: The thick lines show the reflectance $R$
    of the full domain. The dashed line shows the the clear-sky pixels only
    ($R_{\rm clear}$) and the thin solid line shows the cloudy pixels ($R_{\rm
      cloud}$) for the IPA calculation. The lower panel shows corresponding
    differential optical thicknesses, where the lines styles are defined as above.}
  \label{fig:doas_cloud}
\end{figure}

The impact of cirrus clouds on tropospheric NO$_2$ retrievals has been
investigated in a sensitivity study by \citet{vidot2010}. They take into
account the sub-pixel inhomogeneity by a simple independent pixel
approach. Using our Monte Carlo approach we can take into account full
3-dimensional radiative transfer, e.g. the interaction between the cloudy and
the clear-sky part of the domain. Fig.~\ref{fig:doas_cloud} shows an example
where the setup is similar to the study by \citet{vidot2010}. We have taken a
very simple 3D cloud field, the cirrus clouds were modeled as
1$\times$1$\times$1~km$^3$ cubes and arranged as a chess-board, hence the
cloud fraction $c$ is 0.5. The surface albedo is 0.05 and the solar
zenith angle is 30$^\circ$. The optical thickness of the clouds is 3, the
geometrical thickness is 1~km, the cloud base height is 10~km and the
effective radius is 30~$\mu$m where the parameterization by
\citet{baum05a:_bulk, baum05b:_bulk} was used for the cirrus optical
properties. The solar zenith angle is 30$^\circ$ and the wavelength range is
400~nm to 470~nm.  We performed 3D calculation and also used the independent
pixel approximation (IPA) for comparison.  All simulations shown in
Fig.~\ref{fig:doas_cloud} were calculated using MYSTIC with ALIS. 
The reflectance for the IPA simulation is
calculated as the sum of the reflectance of the clear-sky part $R_{\rm clear}$
and the reflectance of the cloudy part $R_{\rm cloud}$ weighted with the cloud
fraction:
\begin{equation}
  R=cR_{\rm cloud} + (1-c)R_{\rm clear}
\end{equation}
In order to speed up the calculations in the presence of clouds, the variance
reduction technique VROOM (\citet{buras2011}) was used. Using VROOM
the simulation 10$^5$ photons are sufficient to obtain an accurate
result with a standard deviation of approximately 0.5\%. 
The 3D calculation using these settings took 1~min~34~s, an IPA calculation
using DISORT with 32 streams takes 58~s.

Fig.~\ref{fig:doas_cloud} shows a part of the spectrum where
we have pronounced features in the differential optical depth. In the
top panel one can see
that for all wavelengths the reflectance in the 3D calculation is smaller than
in the IPA calculation, because photons which are scattered out of the clouds
on the sides have a higher probability of being transmitted to the surface.

The bottom panel of Fig.~\ref{fig:doas_cloud} shows the differential
optical thickness. The difference between IPA and 3D is in this
case about 10\% which will cause an error of some per cent in the
tropospheric NO$_2$ retrievals. Note that this calculation is only an
example to demonstrate the new algorithm to calculate high spectral
resolution spectra using the Monte Carlo method. With different setups
the error on the retrieval can be completely different. 

\section{Conclusions}

We have developed the new method ALIS (absorption lines importance sampling)
that allows to compute polarized radiances in high spectral resolution using
the Monte Carlo method in a very efficient way. We sample random photon paths
at one wavelength. For these random paths we calculate a spectral absorption
weight using the wavelength dependent absorption coefficients
of the model boxes. In order to correct for the wavelength dependence of Rayleigh
scattering an importance sampling method is applied. If necessary the same
method can be applied to correct for the spectral dependence of cloud and
aerosol scattering. The method allows us to calculate radiances for 
many wavelengths at the same time without significantly increasing the
computational cost. ALIS has been implemented in the MYSTIC model
which handles three-dimensional radiative transfer in cloudy atmospheres including
polarization and complex topography.

The new algorithm ALIS has been validated by comparison to the
well-known and well-tested DISORT solver. It has been shown that ALIS
does not produce any bias apart from the statistical noise.  Since all
wavelengths are computed at once, the statistical error is the same at
all wavelength which results mainly in a relative deviation which is
independent of wavelength.  However, for remote sensing applications,
where mostly differential absorption features are of interest, this
deviation does not matter.

Two example applications are shown. First the simulation of polarized
near-infrared spectra over an ocean surface as measured by
e.g.~GOSAT. Here we simulated the Stokes vector with a standard deviation
smaller than 0.05\% for 400 spectral
points in 2 minutes 25 seconds on a single PC. For a standard
deviation of 0.5\% the calculation would be 100 times faster. These short
computation times show that the algorithm has the potential to be used
as a forward model for trace gas retrievals from polarized radiance
measurements, in particular since commonly used discrete ordinate
methods become much slower when they include polarization.
 
The second example is the simulation of the differential optical
thickness from 400~nm to 470~nm which is used to retrieve NO$_2$ from
e.g. SCIAMACHY. Here the computation time for accurate (scalar)
simulations was comparable to DISORT. We performed this calculation also for
an inhomogeneous cloud scene where cirrus clouds are approximated by
simple cubes. We compared the result of the 3D simulation with an
independent pixel calculation and found a difference of about 10~\% in
the differential optical thickness for this example. The calculations
show that ALIS is suitable to study effects of horizontal
inhomogeneity on trace gas retrievals in presence of cirrus clouds.

\section*{Acknowledgement}
We thank Timothy E. Dowling for translating the DISORT code from FORTRAN77 to
C which resulted in great improvements regarding numerical accuracy
and computation time. Furthermore
we thank Jer\^{o}me Vidot for providing NO$_2$ profiles. This work was done
within the project RESINC2 funded by the ``Deutsche
Forschungsgemeinschaft'' (DFG).

\section*{References}
\bibliographystyle{elsarticle-num-names}

\end{document}